\documentclass[11pt,twoside]{article}
\usepackage{asp2010}

\resetcounters

\begin{document}

\title{Wind acceleration in AGB stars: Solid ground \& loose ends}
\author{Susanne H{\"o}fner
\affil{Department of Physics \& Astronomy, Division of Astronomy \& Space Physics, Uppsala University, Box 516, SE--75 120 Uppsala, Sweden}}

\begin{abstract}The winds of cool luminous AGB stars are commonly assumed to be driven by radiative acceleration of dust grains which form in the extended atmospheres produced by pulsation-induced shock waves. The dust particles gain momentum by absorption or scattering of stellar photons, and they drag along the surrounding gas particles through collisions, triggering an outflow. This scenario, here referred to as Pulsation-Enhanced Dust-DRiven Outflow (PEDDRO), has passed a range of critical observational tests as models have developed from empirical and qualitative to increasingly self-consistent and quantitative. A reliable theory of mass loss  is an essential piece in the bigger picture of stellar and galactic chemical evolution, and central for determining the contribution of AGB stars to the dust budget of galaxies. In this review, I discuss the current understanding of wind acceleration and indicate areas where further efforts by theorists and observers are needed.
\end{abstract}

\section{Introduction: General wind properties and simple estimates}\label{s_intro}

Winds with typical velocities of 5--30 km/s are a defining feature of stars on the asymptotic giant branch, shaping their observable properties and sealing their final fate. The heavy mass loss due to the outflows (about $10^{-7}$--$10^{-5} \, M_{\odot}$/yr) strongly affects the evolution of these stars, and the surrounding interstellar medium is enriched with newly produced chemical elements and dust grains. Considering the importance of the phenomenon, it is not surprising that considerable observational and theoretical efforts have been made in order to explore the mechanisms which generate these winds and define their properties. The emerging picture has become more complex 
with new instrumentation and more advanced models but some basic principles remain unchanged and make a good starting point for this review.  

Given the high luminosities of AGB stars, it is natural to assume that stellar radiation is the source of momentum for their winds: Each photon carries momentum corresponding to its energy divided by the speed of light, and the total momentum of the photons emitted by the star per second is given by the stellar luminosity divided by the speed of light ($L_{\ast}/c$). This latter quantity is comparable to, or exceeds, the typical momentum of the gas and dust leaving the star per second, which can be expressed as mass loss rate multiplied by terminal wind velocity ($\dot{M} v_{\infty}$). In the so-called single scattering limit where each stellar photon transfers momentum through one interaction \citep[corresponding to $\dot{M} v_{\infty} = L_{\ast}/c$; see, e.g.,][]{lame99}, a typical star with a luminosity of $L_{\ast}=5000\,L_{\odot}$ should be able to sustain a wind with 10~km/s and a mass loss rate of about $10^{-5}\,M_{\odot}$/yr. 

The argument given above is based on a simple order-of-magnitude estimate and ignores gravity. The efficiency of radiative acceleration depends critically on how much momentum can actually be transferred from photons to matter, or, in other words, on the total radiative cross section per mass of stellar material and the optical depth of the circumstellar envelope. To estimate the amount of opacity required for driving a stellar wind, we consider the relative magnitude of the outwards directed radiative acceleration compared to the inwards directed gravitational acceleration, i.e.,  
\begin{equation}
   \Gamma = \frac{a_{\rm rad}}{a_{\rm grav}} = \frac{\langle\kappa\rangle \, L_{\ast}}{4 \pi \, c \, G M_{\ast}} 
\end{equation}
where $\langle\kappa\rangle$ is the flux mean opacity (total cross section per mass), and $L_{\ast}$ and $M_{\ast}$ denote the luminosity and mass of the star, respectively ($c$ = speed of light, $G$ = gravitation constant). The possible values of $\Gamma$ fall into two distinct regimes, i.e. $\Gamma < 0$ where gravitational attraction dominates and $\Gamma > 0$ where the radiative acceleration exceeds gravity, separated by the special case $\Gamma = 1$ where the two forces are equal. The latter situation corresponds to a critical value of the flux mean opacity, given by
\begin{equation}
   \langle\kappa\rangle _{\rm crit} = \frac{4 \pi \, c \, G M_{\ast}}{L_{\ast}} 
            \approx 2.6 \, \left( \frac{M_{\ast}}{M_{\odot}} \right) \left( \frac{L_{\ast}}{5000 L_{\odot}} \right)^{-1} 
                        \,\, \rm{[cm^2/g]}   \, .
\end{equation}

With this estimate of the values required for driving outflows, we can discuss the relevant sources of opacity. Molecular lines are abundant in the atmospheres of AGB stars, giving them their characteristic visual and infrared spectra. However, the total flux mean opacity of the gas (including both atoms and molecules) is much lower than $\langle\kappa\rangle _{\rm crit}$ and is therefore not a significant factor for radiative acceleration of winds.\footnote{This is in strong contrast to luminous hot stars: their fast winds (with typical velocities of a few 1000 km/s) are presumably driven by atomic line opacity, with Doppler shifts in the wind acceleration zone strongly increasing the effective line width \citep[see, e.g.,][for a detailed discussion]{lame99}.
The latter effect is negligible for radiative acceleration in the much slower AGB star winds.
} 
Dust grains, on the other hand, can be very efficient at absorbing and/or scattering stellar radiation, depending on their chemical composition and size. If dust particles with the right properties are abundant in the outer atmospheric layers, their collective opacity can exceed $\langle\kappa\rangle _{\rm crit}$ and trigger an outflow \citep[see, e.g.,][for a recent compilation of relevant numbers]{BH12}. Ever since characteristic mid-IR features of dust where detected in circumstellar envelopes of cool giants \citep[see, e.g.,][for overviews]{dor10,mol10}, it has therefore been argued that dust grains are responsible for driving the winds of these stars. 

A potential problem with this idea, however, is the fact that solid particles cannot exist (let alone form) at temperatures prevailing in the photospheres of AGB stars: Typical condensation temperatures for common grain types are in the range of 1500--1000~K, or below. A rough estimate for the temperature of dust particles as a function of distance from the center of the star, $T_d (r)$, can be obtained by assuming that (i) the grain temperature is set by radiative equilibrium, (ii) the stellar radiation field can be described with a Planck function corresponding to the effective temperature $T_{\ast}$, geometrically diluted with distance, and (iii) the grain absorption coefficient can be approximated with a power law $\kappa_{\rm abs} \propto \lambda ^{-p}$ in the relevant wavelength region around the stellar flux maximum, where $\lambda$ denotes the wavelength and $p$ is a material-dependent constant \citep[see, e.g.,][for examples of such fits]{BH12}. This leads to 
\begin{equation}\label{eq_td}
   T_d \, (r) \approx T_{\ast} \left( 2r / R_{\ast} \right) ^{- \frac{2}{4+p}}
\end{equation}
where $R_{\ast}$ is the stellar radius \citep[see, e.g.,][for a derivation of this formula]{lame99}. The closest distance at which grains of a given material can form (without evaporating due to radiative heating) is denoted as $R_c$ in the following. It can be estimated by finding the point where the grain temperature is equal to the condensation temperature $T_c$ of the material. According to Eq.(\ref{eq_td}), when setting $T_d \, (R_c) = T_c$, we obtain 
\begin{equation}\label{eq_rc}
   R_c / R_{\ast} \approx \, 0.5 \, \left( T_c / T_{\ast} \right) ^{- \frac{4+p}{2}} 
\, .
\end{equation}
As an example, we consider grains of amorphous carbon which are generally assumed to drive the winds of C-rich AGB stars: in this case, $p \approx 1$ and $T_c \approx 1500\,$K, leading to $R_c \approx$ 2--3$\, R_{\ast}$. As can be inferred from detailed models and observations, these are typical values for the condensation distances of wind-driving dust species (see below). That leaves us with the question of how the dust-free gas can be transported out to several stellar radii (while being under the influence of gravitational attraction), without radiative pressure providing the necessary force. This is where atmospheric shock waves, caused by stellar pulsation or large-scale convection, enter the picture.

\section{The pulsation-enhanced dust-driven outflow scenario}

AGB stars with dusty winds are usually semi-regular or Mira-type long-period variables, as apparent from photometric monitoring. The variations in luminosity are attributed to large-amplitude pulsations, probably combined with large-scale convective motions, which trigger acoustic waves in the stellar surface layers. As the waves propagate outwards through the steep density gradient of the stellar atmosphere they develop into strong radiating shock waves (see, e.g., Freytag, this volume). Similar to water waves breaking on a beach, the shock waves intermittently push parts of the upper atmospheric layers out to a few stellar radii, creating temporary reservoirs of cool dense gas where dust grains may form and grow if the conditions are right. The dust particles gain momentum by absorption and scattering of stellar photons and start moving outwards. Through collisions with the much more numerous gas particles, the dust grains transfer momentum to the gas, dragging it along and thereby triggering an outflow. 

In other words, wind acceleration works like a 2-stage rocket: the first stage (pulsation and/or convection) causes shock waves which send stellar matter on near-ballistic trajectories, temporarily pushing dust-free gas up to a maximum height of a few stellar radii. This provides the starting point for the second stage (radiation pressure on dust) which accelerates the dust-gas mixture beyond the escape velocity. If the second stage fails to kick in, or provides too little acceleration, the gas will fall back towards the stellar surface. In short: both stages are essential for lifting matter out of the stellar potential well and generating an outflow. In the following, I will refer to this scenario as the Pulsation-Enhanced Dust-DRiven Outflow (PEDDRO) scenario. 

\begin{figure}[t]
\begin{center}
\plottwo{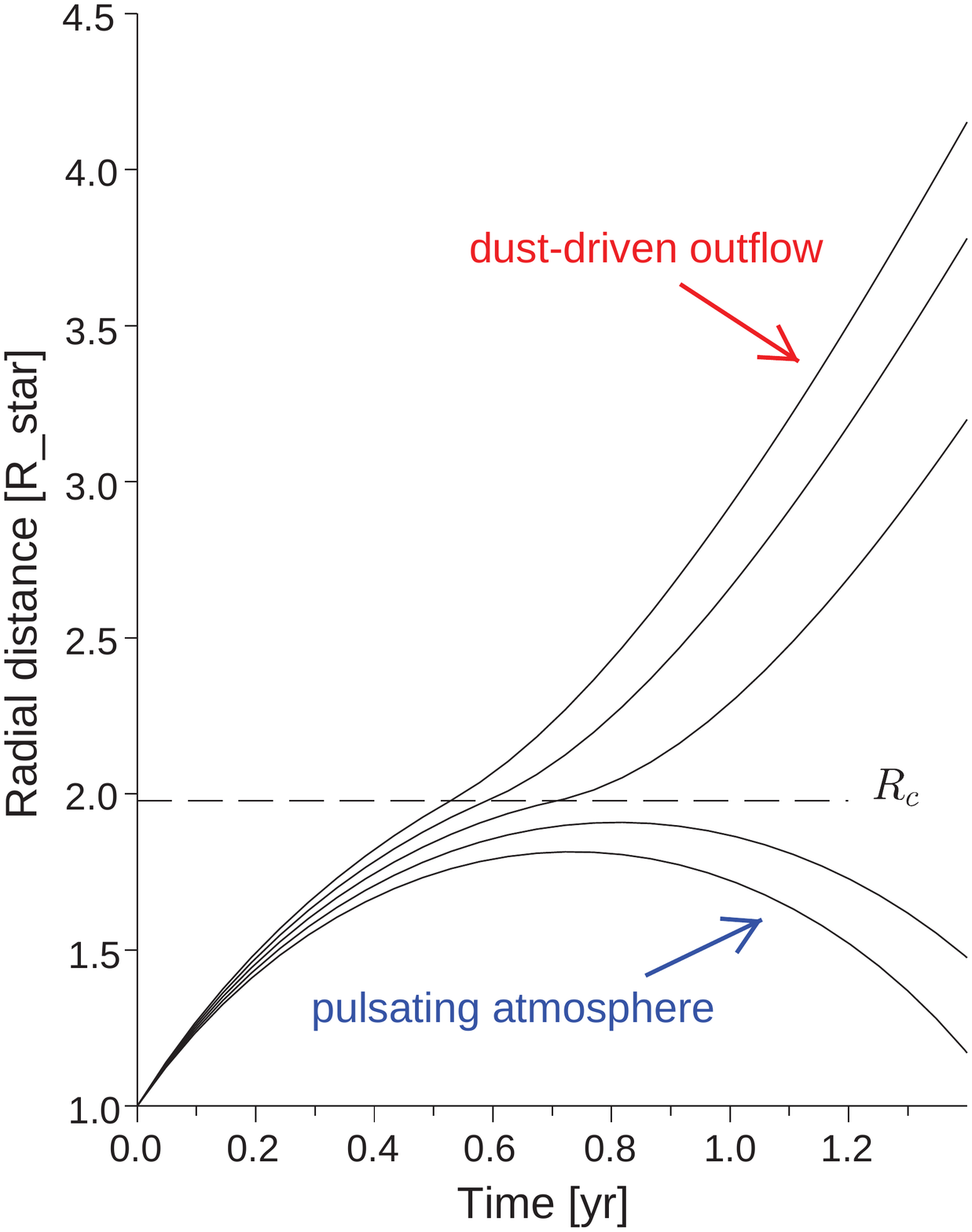}{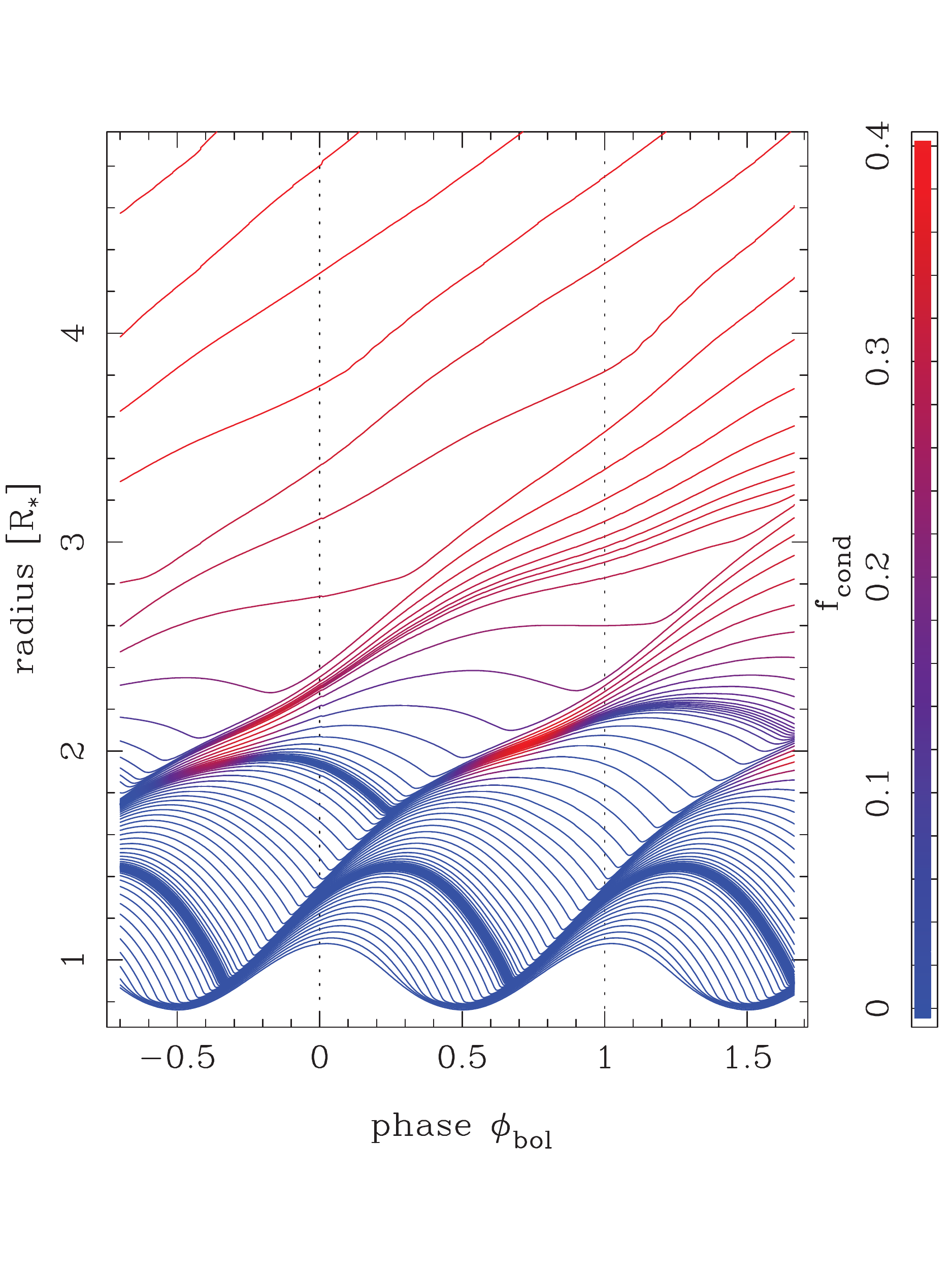}
\end{center}
\caption{Dynamics of the shock-levitated atmosphere and wind acceleration region in the PEDDRO scenario, illustrated by tracking the motions of mass layers. Left panel: a simple toy model \citep[see][]{hoefner09}; Right panel: a detailed radiation-hydrodynamical model \citep[model M in][degree of condensation indicated in color]{nowo10}. Similar stellar parameters are used in both models.} \label{f_toy}
\end{figure}

The dynamics of the atmosphere and wind acceleration region for this scenario is illustrated in Fig.~\ref{f_toy}, tracking the radial motions of mass layers. The left panel gives a schematic picture which corresponds to a simple one-layer toy model, taking only gravity and radiative acceleration into account \citep[see][Sect.~3.1]{hoefner09}. The tracks (which represent different initial velocities)
start where a passing shock pushes the gas outwards. If the initial kinetic energy imparted by the shock is sufficient to reach the condensation distance $R_c$, radiation pressure on dust drives the material outwards with increasing speed ($\Gamma > 1$); otherwise the gas decelerates due to gravity and falls back. The right panel shows the radial movements of mass layers in a detailed radiation-hydrodynamical model with a time-dependent description of dust formation \citep[][model M]{nowo10}. A blue color indicates dust-free gas, red shows the presence of dust. While the overall picture is quite complex, the two basic types of tracks seen in the left panel can also be identified here. 

The time-dependent dynamic behavior of the atmospheres and winds can be studied observationally with high-resolution spectroscopy since Doppler shifts reflecting the gas velocities are imprinted on spectral line profiles. One of the most prominent examples are vibration-rotation lines of CO at near-IR wavelengths which are formed at different depths in the atmosphere and wind, according to their excitation potential. The CO $\Delta \varv = 1$ lines probe the well-established outflows beyond the wind acceleration region, showing more or less pronounces P-Cygni profiles with little temporal variation. The CO $\Delta \varv = 2$ lines form in the dust formation and wind acceleration zone with its complex time-dependent behavior and are difficult to interpret. The CO $\Delta \varv = 3$ lines probe the dust-free inner regions where shock waves dominate the dynamics, showing systematically varying line profiles and periodic line splitting around the luminosity maximum in Mira variables as a shock is passing through the line formation region. Dynamic model atmospheres have been used to derive shock amplitudes from observations of line splitting \citep[e.g.,][]{WWP82,SW2000}. State-of-the-art numerical simulations of the dynamic atmospheres and winds can produce all three types of lines in one consistent model \citep[e.g.,][]{nowo10}.

In addition to spectroscopy, interferometry and imaging have become important tools for studying the structure and dynamics of AGB star atmospheres and wind acceleration zones, providing crucial information about propagating shock waves, dynamics of molecular layers and condensation distances \citep[e.g.,][Wittkowski et al., this volume, Ohnaka, this volume]{tlsw03a,wein04,tevou04,zhao12,karo13}

\section{Quantitative models of mass loss based on the PEDDRO scenario}

While the complex dynamics of AGB star atmospheres and winds is a fascinating topic in itself, it is important not to lose sight of the bigger picture, i.e. a quantitative description of mass loss, which in turn is essential input for stellar and galactic chemical evolution models. A predictive theory of mass loss requires realistic dynamic models that follow the flow of stellar mater all the way from the atmosphere through the wind acceleration region into the circumstellar envelope. The development of self-consistent dynamic models for the PEDDRO scenario has proceeded in 3 major steps over the past decades, investigating the effects of shock waves, non-equilibrium dust formation and frequency-dependent radiative transfer (including detailed molecular and dust opacities), all of which turn out to be crucial for realistic results. 

The pioneering models by \citet{w79} and \citet{b88} represent the first step, studying the role of shock waves for the dynamics of the atmosphere and for wind generation. The efficiency of radiative cooling in the shocks affects atmospheric levitation, and consequently the resulting mass loss rates. More efficient cooling leads to less levitation and lower mass loss rates, since more thermal energy resulting from compression in the shock front is radiated away. In the extreme case of adiabatic shocks (i.e. no radiative cooling), on the other hand, levitation is so efficient that the models show mass loss rates exceeding observations by orders of magnitude \citep[cf.][]{w79}.

Early models of the PEDDRO scenario were built on a simple parameterized description of the dust component, assuming that dust condensation can be described as a function of temperature only \citep[cf.][]{b88}. While this method captures the role of temperature as a threshold for dust formation, it ignores that grain growth rates depend on the ambient gas densities. The second generation of models \citep[e.g.][]{fgs92,hd97,wljhs00,jeong03} includes a detailed, time-dependent treatment of dust formation \citep[][]{GS88,GGS90,GS99}, demonstrating that grain growth proceeds far from equilibrium under the conditions prevailing in atmospheres and winds of AGB stars.  
At distances where temperatures are low enough to allow for dust condensation, gas densities are so low that the time scales of grain growth become comparable to the dynamical timescales (pulsation, convective and ballistic motions), turning the PEDDRO scenario into a race against time: an outflow will form only if condensation is fast and efficient enough that radiation pressure can prevent the shock-levitated gas from falling back towards the star. Furthermore, if an outflow is triggered, rapidly decreasing densities in the wind will quickly quench additional grain growth, usually leading to condensation degrees distinctly below unity. PEDDRO models for M-type AGB stars typically show 20-30\% condensation of Si \citep[e.g.,][]{hoefner08a} and similar values are found for carbon in C-star models \citep[e.g.,][]{matt10} which is in good agreement with recent observational studies (e.g., Khouri et al., Lombaert et al, this volume). This implies that the observed presence of condensible elements in the gas does not exclude the existence of corresponding dust species at the same distance from the star, and vice versa.

With both temperature and density being critical parameters for the onset and efficiency of grain growth, a correct treatment of energy transport in the atmosphere and wind is an essential ingredient of realistic models. The third generation of PEDDRO models \citep[e.g.,][Bladh et al., this volume]{hoefner_etal03,hoefner08a,matt10,erik14} features  frequency-dependent radiative transfer, solved simultaneously with time-dependent gas dynamics and dust formation, taking into account both molecular and dust opacities. While this causes a significant increase in computing time per model, the computational effort is well invested: Realistic gas densities and temperatures in the extended atmosphere require a correct non-gray treatment of opacities, and dust temperatures in models with gray radiative transfer can be incorrect by as much as several hundred Kelvin (as will be discussed in Sect.~\ref{s_td_rc}), with significant consequences for condensation distances, wind acceleration, mass loss rates and synthetic spectra. 

In order to develop a realistic quantitative description of mass loss, models have to be checked against observations. An important tool for measuring wind properties are rotational transitions of molecules at radio wavelengths which have been used extensively to establish both wind velocities and mass loss rates for large samples of AGB stars \citep[e.g.,][]{olof02,gonz03,ramstedt}. Lines of CO, in particular, are useful for probing cool giants of all atmospheric chemistries due to the high abundance and stability of this molecule. State-of-the-art PEDDRO models for both C- and M-type AGB stars give mass loss rates and wind velocities that show reasonable to good agreement with values derived from observations \citep[e.g.,][Bladh et al., this volume]{hoefner08a,matt10,erik14}

\section{Grain temperature and wind-driving dust species}\label{s_td_rc}

A crucial topic that has hardly been mentioned so far is: What types of dust grains are candidates for driving winds? The answer depends on several factors, including radiative cross sections, abundances of the respective chemical elements, and, not the least, condensation distances (which need to fall within the shock-levitated atmosphere in order for the PEDDRO scenario to work). The latter factor is closely related to dust temperatures which, in turn, depend on the optical properties of the grains. 

Due to the strong radiation field around AGB stars, the temperature of dust grains is determined by the balance between heating through absorption of stellar photons and cooling by thermal emission. 
The absorption occurs predominantly at shorter wavelengths than the emission since the star is significantly hotter than the dust particles (i.e., the stellar flux peaks at shorter wavelengths than the dust emission). The relative efficiency of heating and cooling depends on the optical properties of the dust particles in the respective wavelengths regions. Grain materials that have opacities which decrease with wavelength in the near-IR will be more efficient at absorbing than emitting photons, resulting in higher temperatures than for a black body exposed to the same radiation field. The opposite is true for dust particles with opacities that increase with wavelength.

\begin{figure}[t]
\begin{center}
\includegraphics
[width=12cm] {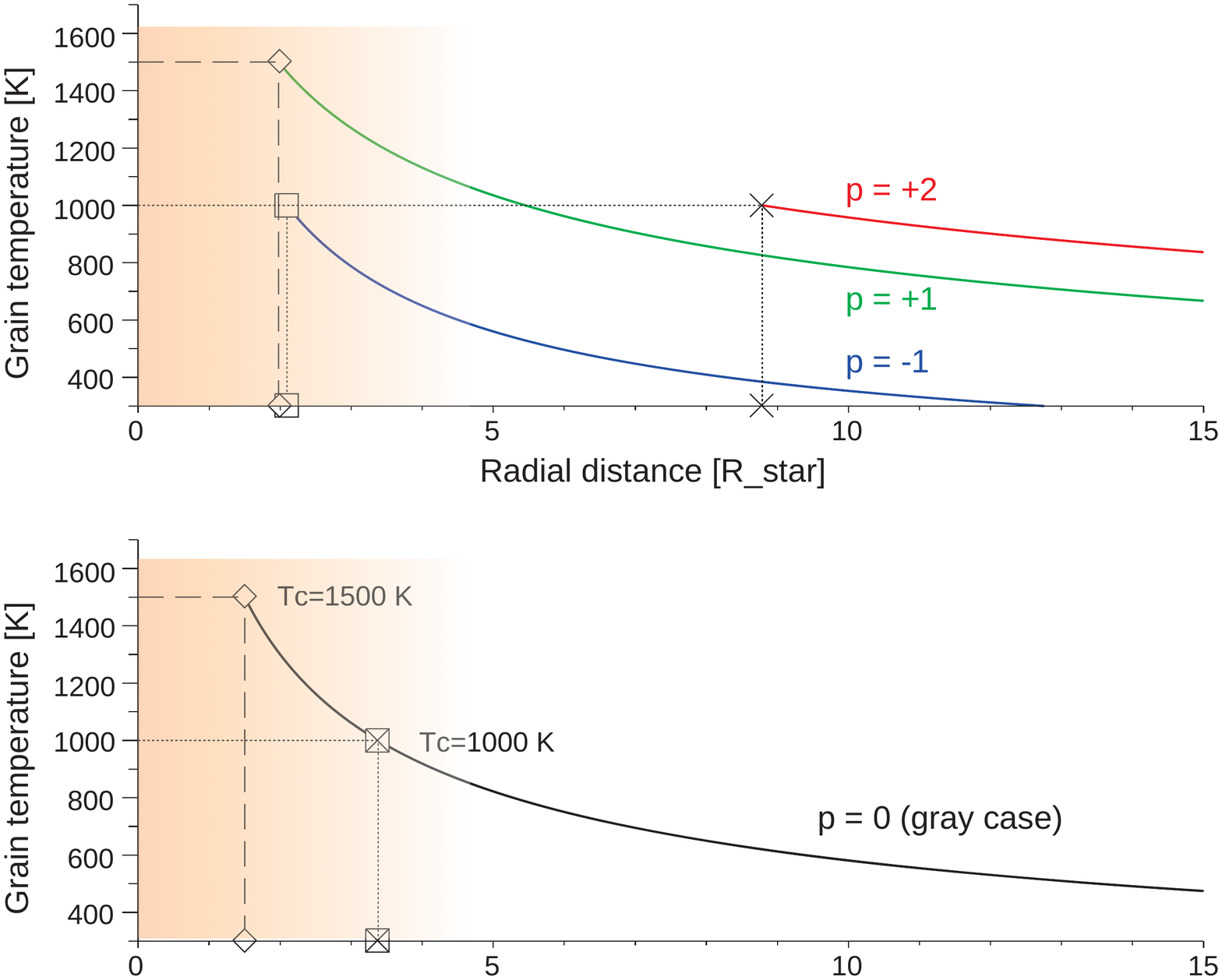}
\end{center}
\caption{Grain temperature as a function of distance from the star ($T_{\ast} = 2600\,$K) according to Eq.~(\ref{eq_td}). The curves in the upper panel represent amorphous carbon ($T_c$ = 1500~K, $p=+1$) and silicates with different Fe/Mg values ($T_c$ = 1000~K, $p=+2$ and $p=-1$). The lower panel shows the gray case (used in some models, leading to incorrect results) for comparison. 
Condensation temperatures and related condensation distances are marked with dashed and dotted lines, and different symbols. 
The shaded area indicates the typical extension of the shock-levitated atmosphere.}\label{f_td}
\end{figure}

This is illustrated in Fig.~\ref{f_td} (upper panel), showing grain temperature as a function of distance from the star according to Eq.~(\ref{eq_td}). In this simple formula, the wavelength dependence of the absorption coefficient in the near-IR is approximated with a power law ($\kappa_{\rm abs} \propto \lambda ^{-p}$, see Sect.~\ref{s_intro}). Comparing the various curves demonstrates that different grain species (characterized by their respective values of $p$) will in general have different temperatures at the same distance from the star. The more positive the value of $p$ (i.e. the steeper the decrease of absorption with wavelength), the hotter the grains will be since radiative heating becomes increasingly more efficient compared to radiative cooling of the grains (greenhouse effect). Consequently, grain species with similar condensation temperatures but different optical properties can have widely different condensation distances. 
This is illustrated in Fig.~\ref{f_td} (upper panel), using typical values for silicate grains: a condensation temperature of $T_c$ =1000~K is combined with $p=+2$ and $p=-1$, representing olivine-type silicates with 50\% and 0\% Fe/Mg content \citep[see, e.g.][]{BH12}, resulting in $R_c \approx 9\,R_{\ast}$ and $R_c \approx 2\,R_{\ast}$, respectively. On the other hand, grain species with different $T_c$ may have comparable $R_c$, depending on their optical properties. In Fig.~\ref{f_td} an example is shown where a higher $T_c$ (1500~K vs. 1000~K) compensates for a higher value of $p$ (+1 vs -1) which leads to a similar value of $R_c$. Note that the numbers in this latter example are representative of amorphous carbon and magnesium silicates, respectively, which are probably the main wind drivers in C- and M-type AGB stars, as will be discussed below.  

For comparison, the lower panel of Fig.~\ref{f_td} shows grain temperatures and condensation distances in the gray case (used in second-generation PEDDRO models and many dust formation models in the literature), which corresponds to setting $p=0$ in Eq.~(\ref{eq_td}). Ignoring the dependence of the optical properties on wavelength leads to the same temperatures for different dust materials (i.e. all are represented by the $p=0$ curve). Consequently, all grain species with similar condensation temperatures $T_c$ are assigned similar condensation distances $R_c$, independent of their optical properties, while different $T_c$ necessarily results in different $R_c$, 
in contrast to the non-gray case. In other words, gray models cannot distinguish properly between grain materials with different optical properties but similar condensation temperatures which is a serious problem for the identification of wind-driving dust species and the prediction of dust yields.

More specifically, the condensation distances of silicate grains are crucial for the much-debated problem of the wind-driving mechanism in AGB stars with C/O $< 1$. Comparing the upper and lower panels of Fig.~\ref{f_td} indicates that gray models \citep[e.g.,][]{jeong03,ferra06} will seriously underestimate the temperature of Fe-bearing silicates ($p=+2$ curve), and consequently their condensation distance ($R_c \approx 3.5\,R_{\ast}$ in the gray case, compared to $R_c \approx 9\,R_{\ast}$ in the non-gray case, in this typical example), leading to the unrealistic conclusion 
that such grains can form close enough to the star to drive the outflows.\footnote{Note that the situation is different for C-type AGB stars: the non-gray effects on the condensation distance of amorphous carbon grains are rather small, as can be seen by comparing the corresponding $R_c$ values in Fig.~\ref{f_td} (about 2 and 1.5 $R_{\ast}$, both well within the shock-levitated atmosphere). Nevertheless, gray wind models are problematic also for these stars: As shown in Fig.~7 of \citet{erik14} the mass loss formula by \citet{wach02} tends to overestimate mass loss rates systematically due to gray atmospheric structures (i.e. too high gas densities) in the underlying wind models.} 
Using a detailed model with non-gray radiative transfer, \citet{woit06b} showed that silicate grains close to the star (i.e. within the shock-levitated atmosphere) have to be essentially Fe-free, and dust particles containing significant amounts of iron can only form at much larger distance. This poses a potential problem for wind acceleration, since Fe-free silicates (Mg$_2$SiO$_4$ or MgSiO$_3$) are quite transparent at visual and near-IR wavelengths, and shock waves will typically not lift stellar matter to distances where Fe-bearing silicates with sufficient absorption cross sections can form. 

A possible solution of this problem is grain size: While the absorption of Fe-free silicate grains is
several orders of magnitude too low for driving an outflow, they can be quite efficient at gaining momentum from scattering of stellar photons, provided that their sizes are comparable to the wavelengths around the stellar flux maximum. \citet{hoefner08a} demonstrated that Mg$_2$SiO$_4$ grains may grow to such sizes (about 0.1 -- 1 $\mu$m) in the dynamical atmospheres, and drive winds with realistic mass loss rates and velocities. \citet{bladh13}
computed synthetic spectra and photometry based on these models, obtaining phase-averaged synthetic (J-K) and (V-K) colors which agree nicely with observations. Even more encouraging, the time-dependent behavior is similar to some well-observed Mira variables, i.e. flat loops in the (J-K) vs. (V-K) diagram with small variation in (J-K) and large variation in (V-K). This is due to variations of molecular features (in particular TiO and H$_2$O) during the pulsation cycle, not an effect of dust opacities. In fact, the good agreement of models with observations in this respect may be taken as a strong indication that the circumstellar envelopes of typical M-type AGB stars are quite transparent at visual and near-IR  wavelengths. 

Observational support for the scenario of winds driven by scattering on Fe-free silicate grains comes from interferometric studies which resolve the spatial structure of the atmosphere and dust formation region. \citet{norris12} measured grain sizes of $0.3\,\mu$m in the immediate vicinity (around 2$\,R_{\ast}$) of several AGB stars, using aperture masking interferometry combined with polarimetry to spatially resolve the light scattered by dust at several near-IR wavelengths. Spectro-interferometry at mid-IR wavelengths (where many dust species show characteristic features due to lattices modes) has been used to derive condensation distances and grain composition in the wind acceleration zone. Signatures of silicate grains are seen at distances of a few stellar radii for some but not all observed stars \citep[e.g.,][]{sacu13,karo13}. These mid-IR studies are complicated by Al-bearing grains which may contribute significantly to the spectra at those wavelengths but have uncertain visual and near-IR optical data which makes consistent modelling difficult.

\section{Conclusions and outlook}

In summary, good progress has been made in understanding wind acceleration in AGB stars in recent years. For carbon stars, PEDDRO models with amorphous carbon grains as wind drivers show improving agreement with observations, regarding, e.g., wind properties \& near-IR photometry 
\citep[e.g.,][]{erik14} or molecular line profile variations \citep[e.g.,][]{nowo10}. For M-type AGB stars, photon scattering on Fe-free silicates seems to be a viable wind acceleration mechanism.  Corresponding PEDDRO models produce realistic wind properties and visual \& near-IR photometry 
\citep[][and this volume]{bladh13}. While both types of models are still far from perfect, a quantitative description of dust-driven mass loss on the AGB seems to be getting within reach. The fact that state-of-the-art PEDDRO models simultaneously reproduce different types of observations rather well may be taken as an indication that most of the basic physical and chemical ingredients are in place. 

Both for the purpose of developing a predictive theory of mass loss (with a minimum of physical parameters) and to improve direct comparisons with observations (not the least spatially resolved data), however, further work is required in a number of areas. Within the framework of current wind models, certain aspects of micro-physics need further study, e.g., optical properties of dust grains (e.g., Zeidler et al., this volume) and non-equilibrium chemistry due to shocks (Cherchneff, Gobrecht et al., this volume). More observational data with high spectral and spatial resolution, for a wide range of wavelengths and, in particular, time series covering at least one pulsation cycle, will be required to better constrain the properties of atmospheric shocks, dust condensation distances and radial wind acceleration profiles. This may also help to clarify the nature of seed particles in AGB stars with C/O$\,<1$. 

On macroscopic scales, one of the most important problems is realistic modelling of the interior dynamics of AGB stars, i.e. pulsation and convection, triggering the atmospheric shock waves that are a crucial part of the PEDDRO scenario. In current wind models, these effects are simulated by rather simplistic variable inner boundary conditions ("piston models") and numerical experiments show that dust formation and wind driving may be rather sensitive to the combined variations in luminosity and velocity (Liljegren et al., this volume). While radial pulsations can be handled in spherical models \citep[e.g.,][]{ireland08}, convection is a phenomenon that is intrinsically three-dimensional, requiring 3D star-in-a-box models \cite[][Freytag, this volume]{frey08}. With more high spatial resolution data and imaging becoming available, 3D structures in atmospheres and winds are becoming more obvious, and the development of 3D star-and-wind-in-a-box models is high on the priority list for the coming years.

\bibliography{hoefner}

\end{document}